\documentclass{aa}
\usepackage{graphicx}
\begin{document}
   \title{Low-mass companions to Hyades stars$^*$}


   \author{E.W. Guenther\inst{1}
          \and
          D.B. Paulson\inst{2}
          \and
          W.D. Cochran\inst{3}
          \and
          J. Patience\inst{4}
          \and
          A.P. Hatzes\inst{1}
          \and
          B. Macintosh\inst{5}
             }

    \offprints{Eike Guenther, \email{guenther@tls-tautenburg.de}\\$*$~
    Partly based on observations obtained at the European Southern
    Observatory at Cerro Paranal, Chile in program 68.C-0063(A), 
    72.C-0288(A) and 072.C-0110(B), and partly based on observations
    obtained at the W.M. Keck Observatory through telescope time allocated to 
    the National Aeronautics and Space Administration through the agency's 
    scientific partnership with the California Institute of Technology and 
    the University of California. The Keck observatory was made possible by the 
    generous financial support of the W.M. Keck Foundation. Also partly 
    based on observations obtained at the Magellan Telescope.}

     \institute{Th\"uringer Landessternwarte Tautenburg,
              Sternwarte 5, D-07778 Tautenburg, Germany
	  \and
             Planetary Systems Branch, 
             Code 693, NASA Goddard Space Flight Center, 
             Greenbelt, MD 20771, USA
          \and
              McDonald Observatory,  
	      University of Texas at Austin, 
	      Austin, TX 78712, USA
          \and
              Division of Physics, Mathematics, and Astronomy
              California Institute of Technology
              Pasadena, CA, USA
          \and
              Institute of Geophysics and Planetary Physics
              Lawrence Livermore National Laboratory 
              7000 East Avenue L-413 
              Livermore, CA 94551, USA
	      }

   \date{Received 10.5.2005; accepted 20.07.2005}

   \abstract{It is now well established that a large fraction of the
    low-mass stars are binaries or higher order multiples. Similarly a
    sizable fraction have giant planets. In contrast to these, the
    situation for brown dwarf companions is complicated: While close
    systems seem to be extremely rare, wide systems are possibly
    more common. In this paper, we present new results on a survey for
    low-mass companions in the Hyades. After measuring precisely the
    radial velocity of 98 Hyades dwarf stars for 5 years, we have
    selected all stars that show low-amplitude long-period trends. With
    AO-observations of these 14 stars we found companion candidates
    around nine of them, where one star has two companions. The two
    companions of HIP 16548 have masses between 0.07 to 0.08
    $M_{\odot}$, and are thus either brown dwarfs or very low mass
    stars. In the case of HAN 172 we found a companion with a mass between
    0.08 to 0.10 $M_{\odot}$, which is again between a star and a brown
    dwarf. The other seven stars all have stellar companions. In two
    additional cases, the RV-variations are presumably caused by stellar
    activity, and in another case the companion could be a short-period
    binary.  The images of the remaining two stars are slightly
    elongated, which might imply that even these are binaries. 
    Because at least 12 of the 14 stars showing low-amplitude RV trends
    turn out to have companions with a mass $\geq 70$ $M_{\rm Jupiter}$,
    or are just active, we finally estimate the number of companions
    with masses between 10 $M_{\rm Jupiter}$ and 70 $M_{\rm Jupiter}$
    within 8 AU of the host stars in the Hyades as $\leq$ 2\%.

    } \maketitle


\section{Introduction}

It is now well established that most of the solar-like stars are
binaries (Abt \& Levy \cite{abt76}; Duquennoy \& Mayor
\cite{duquennoy91}). A detailed analysis shows that $13\pm3$\% of
the G stars, and 8.1\% of the M stars are binaries with separation of 3
AU, or less (Mazeh et al. \cite{mazeh92}; Fischer \& Marcy
\cite{fischer92}). The frequency of planets with a mass of $m\,\rm
sin\,i$ $\geq$ $0.3\,M_{\rm Jupiter}$ orbiting at distances $\leq 5$ AU
is about 9\% (Lineweaver \& Grether \cite{lineweaver03}). It is quite
surprising that brown dwarfs are very rare as close companions to normal
stars.  The lack of brown dwarfs as companions is thus often referred as
the brown dwarf desert.  Marcy et al. (\cite{marcy03}) estimate from
their radial velocity (RV) survey that the frequency of brown dwarfs
with 3 AU of the host stars is only $0.5\pm0.2\%$, and thus much smaller
than the frequency planets, or the frequency binaries. Studies by Zucker
\& Mazeh (\cite{zucker01}) show that the frequency of close companions
drops off very sharply for masses higher than 10 $M_{\rm Jupiter}$,
although they suspect there is still a higher mass tail that extends up
to probably 20 $M_{\rm Jupiter}$. From the currently known 161
``planets'' 8 have an $\rm m\,sin\,i$ between 10 and 18
$\,M_{Jupiter}$ 
\footnote{as compiled from J. Schneiders
``Extrasolar Planets Encyclopaedia'', http://www.obspm.fr/encycl/catalog.html'')}.
There are only very few
brown dwarfs known that orbit normal stars at distances $\leq 3$ AU
(Zucker \& Mazeh \cite{zucker00}; Udry et al.  \cite{udry02}; Endl et
al.  \cite{endl04}).

In the case of wide pairs consisting of a brown dwarf and a star, the
situation is more complicated. These wide companions (e.g. $d \geq 50$
AU) are usually detected by means of direct imaging.  Direct imaging
campaigns probably have turned up a number of brown dwarfs orbiting
normal stars (Gliese 229B: Nakajima et al. \cite{nakajima95}; TWA-5\,B:
Lowrance et al. \cite{lowrance99}; Neuh\"auser et
al. \cite{neuhaeuser00}; HR~7329\,B: Lowrance et al. \cite{lowrance00};
Guenther et al.  \cite{guenther01}; Gliese 417B and Gliese 584C:
Kirkpatrick et al. \cite{kirkpatrick01}; HR~7672: Liu et
al. \cite{liu02}; HD~130948 (two brown dwarfs): Goto et
al. \cite{goto02}, Potter et al. \cite{potter02}; LHS~2397aB: Freed et
al.  \cite{freed03}, Masciadri et al. \cite{masciadri03}; GSC
08047-00232: Neuh\"auser \& Guenther \cite{neuhaeuser04};
$\epsilon$\,Indi\,B: McCaughrean et al. \cite{mccaughrean04}; G124-62
Seifahrt et al. \cite{seifahrt04}). Additionally to these detections
there are three objects that are either low-mass brown dwarfs, or even
giant planets (Neuh\"auser et al. \cite{neuhaeuser05}; Chauvin et
al. \cite{chauvin05a}; Chauvin et al. \cite{chauvin05b}). The result of
all search programs for objects in TWA-Hydra, Tucanae, Horologium and
the $\beta$ Pic region is that the frequency of of brown dwarfs at
distances larger than 50 AU is $6\pm4\%$ (Neuh\"auser et
al. \cite{neuhaeuser03}). This result implies that the frequency of wide
binaries consisting of a brown dwarf and a star is much higher than that
of close binaries. One possible explanation for these results may simply
be the difference of the methods used for detecting short and long
period systems. In the case of the long-period systems, the masses of
these objects have been determined from absolute brightnesses by
evolutionary tracks. If for some reason the evolutionary tracks are off
by a certain factor, a number of these brown dwarfs would be
stars. There are in fact notable differences between evolutionary tracks
published by different authors, especially for objects of lower mass,
and young age (Wuchterl \& Tscharnuter \cite{wuchterl03}; Stern
\cite{stern94}; Burrows et al. \cite{burrows97}; Malkov et al.
\cite{malkov98}; Chabrier et al. \cite{chabrier00}). A recent
determination of the dynamical mass of \object{AB Dor C} indicates that
the true mass is a factor of two higher than the mass derived from
evolutionary tracks, converting this brown dwarf into a star (Close et
al. \cite{close05}).  On the other hand, in the case of binary brown
dwarf \object{GJ 569 Bab} the true mass and the mass derived from
evolutionary tracks agree reasonably well (Zapatero Osorio et al.
\cite{zapatero04}).  Possible explanations for the discrepancy could be
that \object{AB Dor C} is a binary, or that \object{AB Dor} has an
age of only 50 Myr, whereas \object{GJ 569 Bab} one between 120 and 1000
Myr. A study of old, isolated G, K, and M stars indicate that there is
also a brown dwarf desert for long period systems, as McCarthy \&
Zuckerman (\cite{mccarthy04}) derive a frequency of $1\pm1\%$ for brown
dwarfs orbiting stars between 75 and 300 AU.

We can thus summarize that there is a general agreement that the
frequency of brown dwarfs orbiting stars at distances of $\leq 3$~AU is
about $\leq 0.5\%$. For brown dwarfs at larger distances, there are two
somewhat contradicting results: one is that the frequency is
$1\pm1\%$, the other that it is $6\pm4\%$.

The aim of this work is to shed more light on the brown dwarf desert
problem by a novel approach: Brown dwarfs out to a distance of
about 8 AU can be detected by means of radial velocity measurements,
because these would show up as trends in the radial velocity data, and
can thus be detected in this way. The only problem is that a stellar
companion at a larger distance will also induce a trend of the
RV-data. The trick is that wide binary stars can easily be detected by
means of adaptive optics (AO) imaging. Objects that show low-amplitude
trends in the RV-data, and are not binary stars are thus brown dwarf
candidates, or even giant planets.  In other words, if all stars that
show low-amplitude trends in the RV-data turn out to be binary stars, we
can conclude that it is highly unlikely that any of the stars monitored
has a brown dwarf companion orbiting the primary within 8 AU. The
advantage of this method is that the selection of the objects is based
on RV-measurements, and thus all brown dwarfs within 8 AU of the host
stars are detected, unless the orbit is very unfavorable.

For this project we select a cluster of stars: the Hyades. This
has three advantages.  First of all, a cluster is a homogeneous sample,
as all stars have the same age, metallicity, and distance. Secondly, the
Hyades are sufficiently young ($625\pm50$ Myrs, Perryman
\cite{perryman98}; or 650 Myrs Lebreton et al. \cite{lebreton01}) and
close enough ($46.34\pm0.27$~pc, Perryman \cite{perryman98}) so that not
only stellar companions can be detected but also brown dwarfs.  The
third reason is that this sample was neither studied by Neuh\"auser et
al. (\cite{neuhaeuser03}), nor by McCarthy \& Zuckerman
(\cite{mccarthy04}) and thus serves as a independent investigation that
can be compared with these results.  Additionally, the Hyades are
metal-rich ([Fe]/[H] = +0.17$\pm$0.06: Boesgaard \& Budge
\cite{boesgaard88}; [Fe]/[H] = +0.12$\pm$0.03: Cayrel et al
\cite{cayrel85}; [Fe]/[H] = +0.13$\pm$0.01: Paulson et
al. \cite{paulson03}). Whether the frequency of brown dwarf companions
is related to the abundance or not is not known but it is known that the
frequency of planets is higher for metal rich stars (Santos et
al. \cite{santos04}).  Also in this respect it is interesting to
compare the frequency of brown dwarf companions of Hyades stars with
that of normal stars.


\section{Observations}

\subsection{Radial velocity: The input sample}

The RV measurements were carried out with the HIRES spectrograph on the
10-m-Keck telescope over a period of about 5 years (see Cochran et
al. \cite{cochran02}; Paulson et al. \cite{paulson02} and Paulson et
al. \cite{paulson04} for details).  After measuring the RVs of 98 Hyades
dwarf stars for 5 years, we identified 13 stars that show long-period,
linear trends. The objects selected have trends with a velocity gradient
($\Delta RV$) between 5 and 100 $m/s\,yr^{-1}$.  At the elongation, the
gradient for a Hyades-star in a circular orbit is $\Delta RV =
6.6\,(m_{comp}/[80 m_{Jupiter}])\, (
sep)^{-2}\,[m\,s^{-1}\,yr^{-1}] $ (with $m_{comp}$ the mass of the
companion, and $sep$ the separation in arcsec).  An object showing a
trend of this order thus is most likely either a wide system consisting
of two stars, or a system consisting of a star and a substellar-mass
companion. By imaging these objects with an AO-system, we can
distinguish between stellar- and substellar- mass companions. Combining
RV-data with AO-imaging of relatively young objects also is a good way
of finding suitable targets for attempts to directly image planets with
future instruments like CHEOPS (Claudi et al. \cite{claudi04}).
Table\,\ref{tab:objects} gives a short overview of the objects chosen.

\begin{table}
\caption{The stars}
\begin{tabular}{lllll}
\hline \hline
            & spec  & $m_V$ & RA (2000.0) & Dec (2000.0) \\  
\hline
HD\,28099   & G2V   & 8.1  & 04 26 40.1 & +16 44 49 \\
HD\,30505   &       & 9.0  & 04 49 03.5 & +18 38 28 \\
HIP\,16908  & K2V   & 9.4  & 03 37 35.0 & +21 20 35 \\
HD\,286589  & K5V   & 10.8 & 04 14 51.9 & +13 03 18 \\
LP\,415-176 & K2V   & 11.6 & 04 31 44.5 & +15 37 46 \\
HIP\,16548  & M0V   & 12.0 & 03 33 05.3 & +04 57 29 \\
V1102\,Tau  & M1V   & 12.2 & 04 28 28.7 & +17 41 45 \\
HAN\,513    & M0.5V & 12.3 & 04 31 43.3 & +15 02 29 \\
HAN\,172    & M1V   & 12.5 & 04 17 47   & +13 39 42 \\
HIP\,15720  & K7V   & 8.5  & 03 22 28.1 & +27 09 22 \\
BD+08\,642  & K5V   & 10.1 & 04 09 49.4 & +09 18 20 \\
BD+16\,630  &       &      & 04 36 00   & +16 32 30 \\ 
J\,332      & K5V   & 11.5 & 04 47 35   & +14 53 24 \\
LP\,415-378 & M0V   & 12.4 & 04 29 12   & +15 16 12 \\
\hline\hline
\end{tabular}
\label{tab:objects}
\end{table}

\subsection{Direct imaging : NACO}

Most of the direct images were carried with the NAOS-CONICA-system
(Nasmyth Adaptive Optics System - High-Resolution Near IR Camera) at the
VLT-UT\,4 telescope Yepun of the European Southern Observatory at Cerro
Paranal, Chile in program 72.C-0288. We used the S27-camera, which gives
a field-of-view of 28x28 arcsec. An image scale of $27.07\pm0.02$
mas/pixel was recently derived by Neuh\"auer et
al. (\cite{neuhaeuser05}) using Hipparcos stars. The new value of the
image-tilt is $0.14\pm0.10^o$ (towards the east).  Our original plan
was to observe all objects with and without a semi-transparent
coronographic mask but not all observing blocks could be carried out.

\begin{figure}[h]
\includegraphics[width=0.35\textwidth, angle=270]{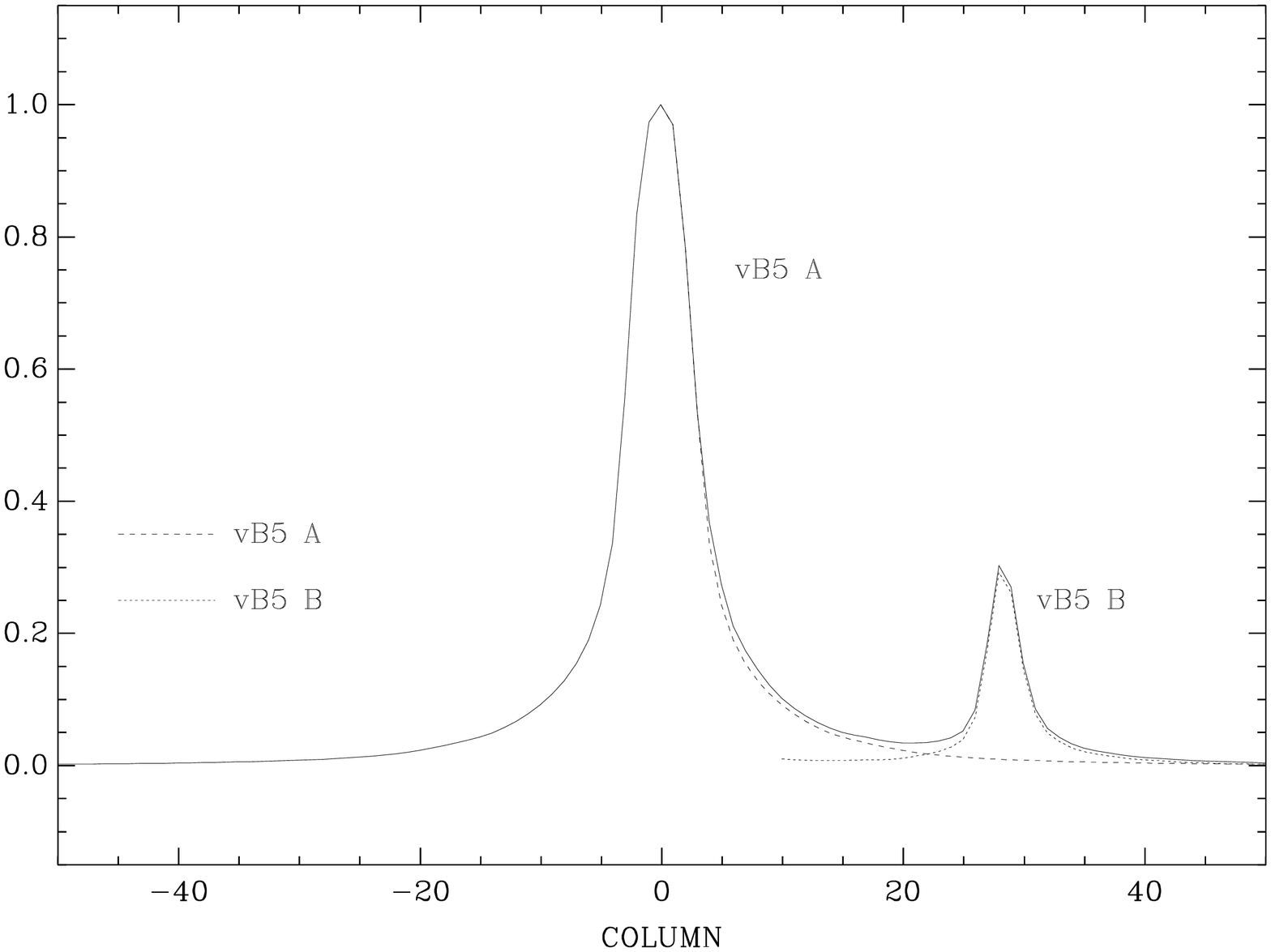}
\caption{The figure shows how the fluxes of the primary and
secondary are measured individually. The full-line shows a cut through
the image of HIP16908\,A. The two dashed lines are: 1.) The profile of
the A-component after subtracting the B-component 2.) The B-component
after subtracting the A-component.}
\label{HIP16908_phot}
\end{figure}

The semi-transparent mask has a projected diameter of 0.7 arcsec, and
dims the light of the central star by about 10 magnitudes. This mask is
thus very useful for detecting faint objects at separations larger than
0.35 arcsec, and bright objects closer to the star.  It also allows us
to measure the position angles and separations if a companion is found.
According to the evolutionary tracks calculated by Burrows et al.
(\cite{burrows97}) objects of 20 $M_{Jupiter}$ have nearly the same
brightness in the J, and in the {\it K-band}.  Because with NACO a
similar limiting magnitude is reached easier in in the {\it J-band} than
in the {\it K-band}, we used {\it J-band} filter with a central
wavelength 1.265 $\mu m$ for these observations.  The exposure-time was
set in order to reach a 3 $\sigma$ limiting magnitude between J=22.5 mag
and J=23 mag. According to Masciadri et al. (\cite{masciardi05}) NACO
allows to detect an object 12 mag fainter than the primary at the 5
$\sigma$-level under average observing conditions. The limit thus is 20
to 22 mag at a separation of one arcsec, which corresponds to objects
with a mass of 10 to 20 $M_{Jupiter}$ (using Burrows et
al. \cite{burrows97}). We thus can not only detect companions stars but
also the brown dwarfs themselves at distances larger than about 16 AU.
Table\,\ref{tab:imaging_naco} gives an overview of the data taken, where
``coro'' or ``direct'' indicates whether the semi-transparent
coronographic mask was used, or not used. ``SR'' is a conservative
estimate of the Strehl ratio. The shortest possible exposure time
(0.3447s) was used for ``direct'' observations in order to minimize, or
avoid the saturation of the primary stars. Because of the corresponding
increase of the read-out noise, the on-source observation time had to
increased in order to reach about the same limiting-magnitude, as in the
coronographic images.  The data was taken in service mode in ten
different nights. Three of the observing nights (16 Dec 2003, 5 Nov
2003, 19 Feb 2004; MJD=52928, 52948, 53054) were categorized as dry
(relative humidity below 10\%), and five (25 Dec 2003, 12 Jan 2004, 26
Feb 2004, 7 Mar 2004, 12 Mar 2004; MJD=52998, 53016, 53061, 53071,
53076) as normal (relative humidity 10 to 25\%).  Unfortunately
\object{J332} and \object{HAN\,172} (9 Feb 2004, 10 Feb 2004; MJD=53044,
53045) were observed during wet conditions (relative humidity 56\%, and
62\%). These two nights were also partly non-photometric. The accuracy
of the photometry of these two stars is thus limited. For the other
nights, we could use the fluxes of the primaries (when not saturated,
and when taken without the mask), as well as observations of ESO's
flux-standard stars taken at the beginning of each night, to
flux-calibrate the images. In the case of stars observed with the mask,
only the standard stars could be used and thus, the photometric errors
are a bit higher.  Fig.\,\ref{HIP16908_phot} shows a typical example how
the fluxes of the two components of a binary are measured individually.
The errors of the separation and position angle for the binaries given
in Table\,\ref{tab:results} were determined from the variance of the
individual exposures and the systematic error of the determination of
the image scale and image tilt given above.

\begin{table}
\caption{Observations : NACO}
\begin{tabular}{lrccr}
\hline \hline
            & date & filter, SR & set-up & exposure-\\
            & & & & time \\
\hline
HD\,30505   & 7 Mar 2004  & J,12.2\% & coro   & 800s  \\
HIP\,16908  & 5 Nov 2003  & J,17.9\% & direct & 1586s \\
HD\,286589  & 25 Dec 2003 & J,11.7\% & direct & 1586s \\
LP\,415-176 & 5 Nov 2003  & J,11.5\% & direct & 2061s \\
HIP\,16548  & 16 Dec 2003 & J,12.0\% & coro   & 960s  \\ 
V1102\,Tau  & 12 Mar 2004 & J,11.9\% & coro   & 960s  \\
HAN\,513    & 26 Feb 2004 & J,11.6\% & direct & 1586s \\ 
HAN\,172    & 10 Feb 2004 & J,11.5\% & coro   & 960s  \\ 
HIP\,15720  & 5 Nov 2003  & J,17.9\% & coro   & 960s  \\
BD+08\,642  & 5 Nov 2003  & J,12.0\% & coro   & 960s  \\ 
BD+16\,630  & 19 Feb 2004 & J,11.7\% & coro   & 960s  \\ 
J\,332      & 9 Feb 2004  & J,11.6\% & coro   & 960s  \\
LP\,415-378 & 12 Jan 2004 & J,11.7\% & direct & 1586s \\ 
\hline\hline
\end{tabular}
\label{tab:imaging_naco}
\end{table}

\subsection{Direct imaging : Keck II}

In addition to the NACO AO-observations, \object{HD\,30505},
\object{HD\,286589} and \object{HD\,28099} were observed with the Keck
II telescope on Dec 8 2003.  We used the narrow field (pixel scale
$0.00993'' \pm 0.00005''$; Ghez et al. (\cite{ghez04}) NIRC2 facility
camera \footnote{see alamoana.keck.Hawaii.edu/inst/nirc2 for details}
which accepts the corrected beam from the adaptive AO-system (Wizinowich
et al. \cite{wizinowich00}).  The weather conditions throughout the
night were marked by increasing amounts of clouds which impact the
AO-system performance. The first two targets observed -
\object{HD\,28099} and \object{HD\,30505} - could be observed with the
AO-system wavefront sensor operating at the fastest rate, however,
extinction from clouds forced the last target - \object{HD\,286589} - to
be observed at a reduced update rate. The lower sampling rate resulted in
a slightly broader FWHM for this source. In contrast to NACO, objects of
lower mass can be detected with NIRC2 if the K' band is used. This is
because the Strehl ratio improves at longer wavelength. The other
reason for using the K' band is that for those objects, which were
observed with both telescope, we obtain color information.  Exposure
times were chosen to avoid saturation on the bright primary star.  The
exposures where continued, even when a companion became visible in order
to search for possible additional, fainter companions.

The exposure time listed in Table\,\ref{tab:imaging_keck} is the sum of
the eight individual exposures taken at four different array
positions. For \object{HD\,28099} and \object{HD\,30505} an AO-rate of
672 Hz was used, for \object{HD\,286589} we used an AO-rate of 210 Hz.
The final positions, angles, and magnitude differences listed in
Table\,\ref{tab:results} are determined from a shifted and combined
image created from all exposures.  The uncertainties of the measurement
of the positional angle and the distance are derived from the systematic
error of the pixel scale ($\pm 0."00005$; Ghez et al. \cite{ghez04}),
and position angle ($\pm 0.33^o$; Lu \cite{lu04}) plus the standard
deviation of the measurement as derived from the measurement of each
individual image.  Because the conditions were not photometric, the
measured K' magnitude difference and the 2MASS Ks primary magnitude were
used to estimate the secondary magnitude; no transformations were
applied.

\begin{table}
\caption{Observations: Keck}
\begin{tabular}{lcclcc}
\hline \hline
Name       & date   & Filter, SR & set-up & exposure-\\
           & & & & time \\
\hline
HD\,28099  & 8 Dec 2003 & H2(1-0),37\% & direct & 80s \\
HD\,30505  & 8 Dec 2003 &      K',41\% & direct & 64s \\
HD\,286589 & 8 Dec 2003 & H2(1-0),27\%,& direct & 80s \\
\hline\hline
\end{tabular}
\label{tab:imaging_keck}
\end{table}

\section{Results}


\begin{figure}[h]
\includegraphics[width=0.35\textwidth, angle=270]{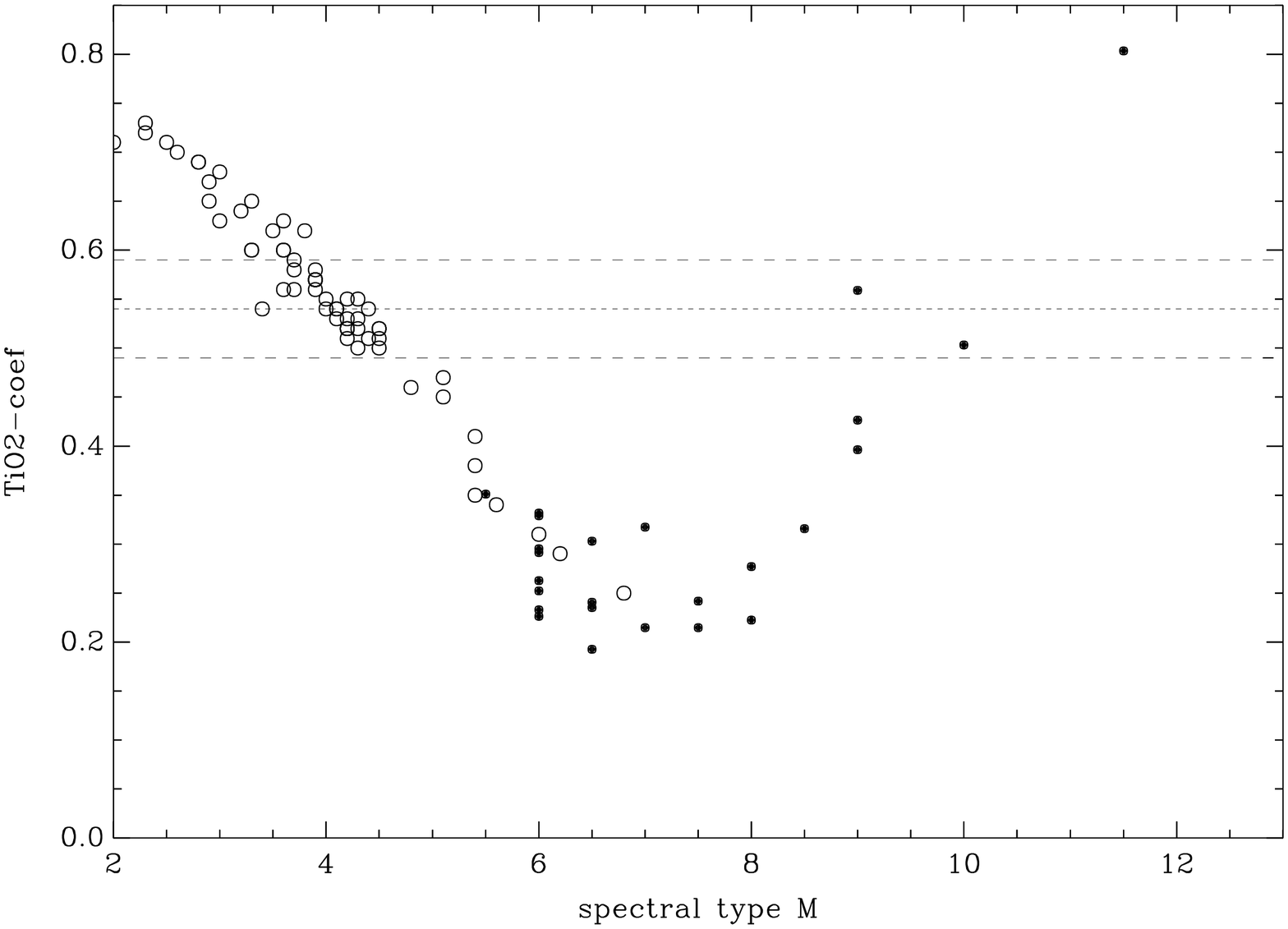}
\caption{The dashed line shows the TiO2-coefficient together with the
errors. The circles are values obtained for M stars from Reid et
al. (\cite{reid95}) and Hawley et al. (\cite{hawley96}) for Hyades
stars, and the filled dots for very low-mass stars from Guenther \&
Wuchterl (\cite{guenther03}). The spectral type thus is between M9V and
L0V, in good agreement with the brightness of the object in the {\it J-band}.}
\label{TiO2-spec-type}
\end{figure}

\begin{figure}[h]
\includegraphics[width=0.35\textwidth, angle=270]{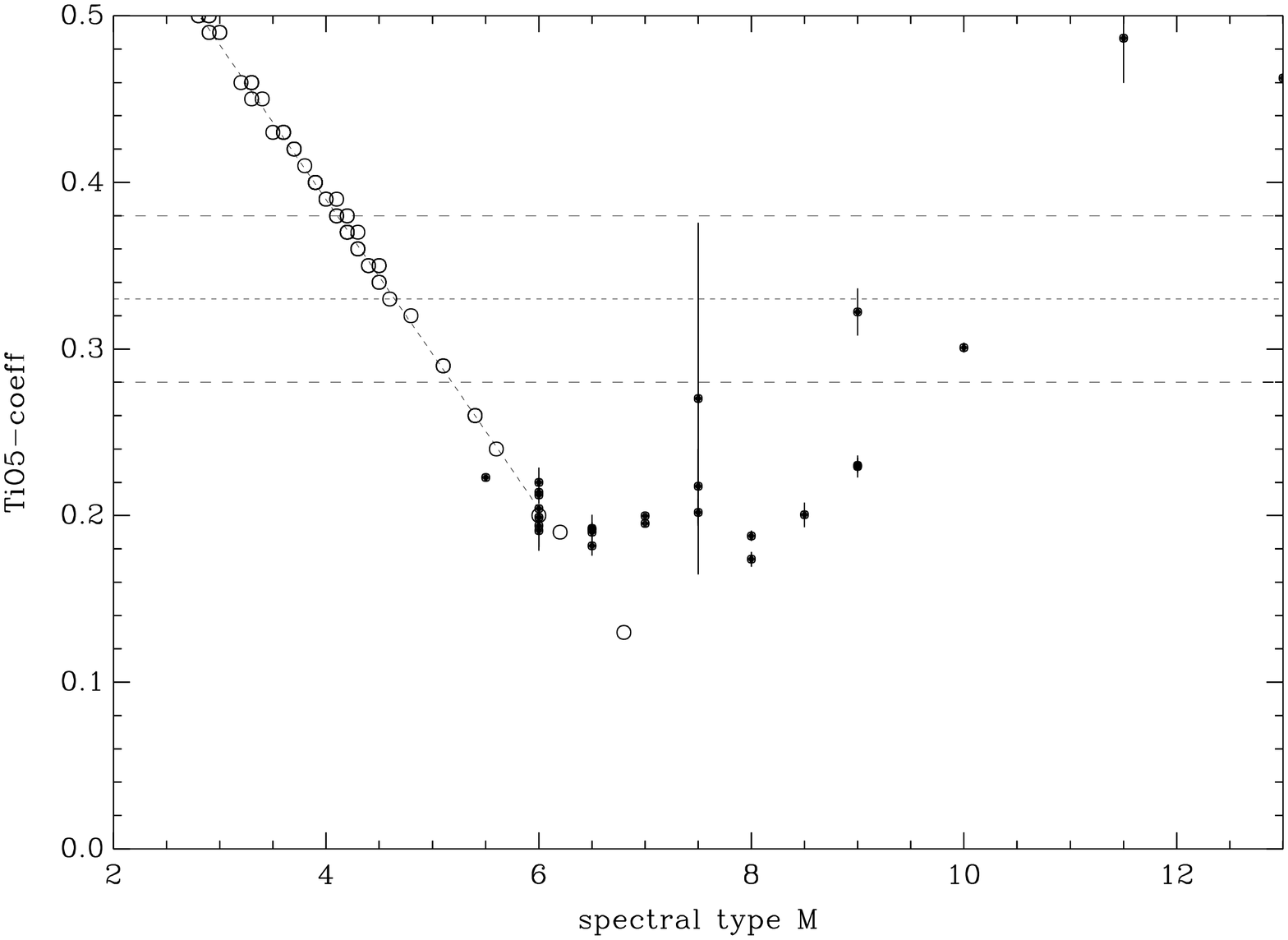}
\caption{Same as Fig.\,\ref{TiO2-spec-type} but for the
TiO5-coefficient. We find again a possible solution for the spectral
type of M9V and L0V.}
\label{TiO5-spec-type}
\end{figure}

\subsection{HIP\,16548}

\object{HIP\,16548} is an M0V star. We found two companion candidates
west of the star (Fig.\,\ref{HIP16548-NaCo}). The separations to the
primary are $2.77\pm0.02$ and $2.78\pm0.02$ arcsec, respectively. The
separation between the two companion-candidates is $0.30\pm0.02$ arcsec
(Table\,\ref{tab:results}).  If true companions, the separation between
the primary and the companions would be 130 AU, and the separation
between the two 14 AU. The orbital periods would thus be of the order of
2000 and 100 years, respectively. The companion-candidates are
$13.8\pm0.3$ and $14.0\pm0.3$ mag in the {\it J-band}. This would
correspond to absolute magnitudes $M_{\rm J}$ of 10.5 and 10.7 mag,
which implies masses below 0.1 $M_{\odot}$.

We took a spectrum with the MIKE (Magellan Inamori Kyocera Echelle)
spectrograph on the Magellan 6.5m telescope. The resolution is $\lambda
\slash \Delta \lambda$ is 30,000 at 7000 \AA\ with a slit width of 0.7
arcsec. Thus, both components were in the slit.  The integration time
was one hour, which gave a S/N of about 20. Following Reid et
al. (\cite{reid95}) and Hawley et al. (\cite{hawley96}), the spectral
types of M stars can be determined by measuring the height of the
TiO-bandheads. The TiO2 and TiO5-coefficients are the ratios of
the fluxes TiO2=F[7043-7046 \AA ]/F[7058-7061 \AA ] and TiO5=F[7042-7046
\AA ]/F[7126-7135 \AA ].  Figs. \ref{TiO2-spec-type} and
\ref{TiO5-spec-type} show the TiO2 and TiO5-coefficients versus the
spectral type.  As can be seen in the figures, both coefficients
decrease up to a spectral-type of M7V and then increase again. For a
given coefficient, there could be two possible spectral types, and hence
masses.  Reid et al.  (\cite{reid95}) and Hawley et
al. (\cite{hawley96}) both limit their analysis to stars earlier than
M7V.  We use our own library of spectra (Guenther \& Wuchterl
\cite{guenther03}) to expand the method up to spectral type L. The
dashed lines in Fig.\,\ref{TiO2-spec-type} and
Fig.\,\ref{TiO5-spec-type} give the TiO2 and TiO5-values obtained for
HIP\,16548\,B and C. The TiO-coefficients give as possible solutions a
spectral-type M9V to L0V, and M4V to M5V. In order to find out which of
two possibilities is the correct one, we additionally used the
PC3-coefficients from Mart\'\i n et al. (\cite{martin96}). From the
PC3-coefficients PC3=F[8230-8270 \AA ]/F[7540-7580 \AA ]; SpT =
-6.685 + 11.715 $\times$ (PC3) - 2.024 $\times$ $(PC3)^2$ ), we
obtained a spectral type M8V to M9V.

\begin{figure}[h]
\includegraphics[width=0.35\textwidth, angle=270]{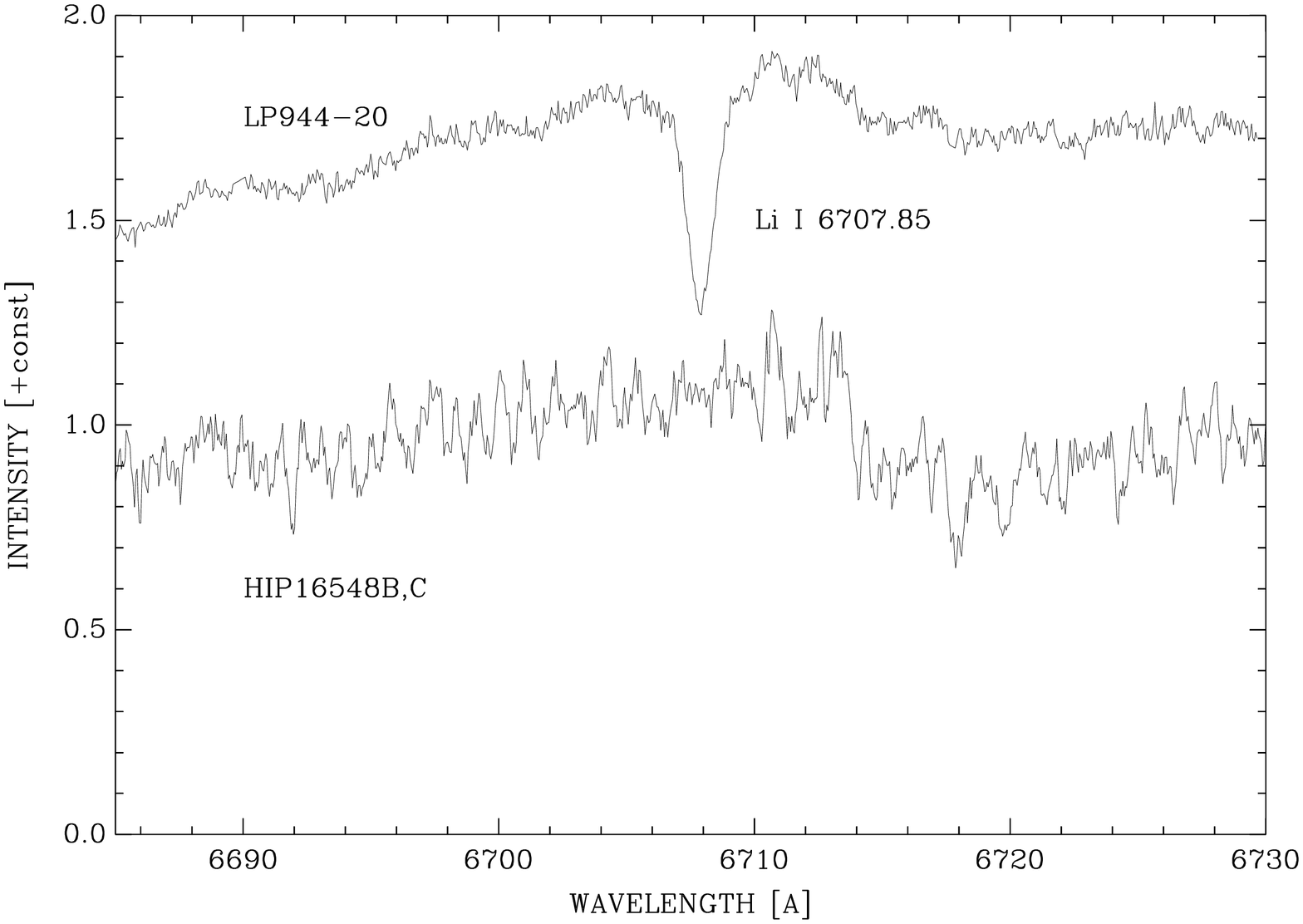}
\caption{The figure shows the Li I 6708 \AA \, line in the
spectrum of the brown dwarf LP944-20 which has about the
same age and the same spectral type as HIP\,16548\,B,C. This line is not
seen in the spectrum of HIP\,16548\,B,C, which is shown below.}
\label{Li6708_spec}
\end{figure}

The fact that we now know that the objects have a spectral type of M8V
to M9V allows us to test the quality of the photometry.  According to
Dahn et al. \cite{dahn02} an objects with a spectral type in the range
between M8V and M9V have an $M_J$ in the range between 10.5 and 11.8
mag.  For \object{HIP\,16548\,A} and \object{HIP\,16548\,B} we derive
$M_J=10.5\pm0.3$ and $M_J=10.8\pm0.3$ using the distance of the Hyades
and the values given in Table\,\ref{tab:results}. The agreement between
the spectroscopic and photometric determination of the spectral type
thus is excellent.  The objects have masses between 0.07 to 0.08
$M_{\odot}$ (Chabrier et al. \cite{chabrier00}), which puts the objects
very close to the brown dwarf boarder. We do not, however, see the
Li\,I\,6708 line, so it is unclear as to whether the objects are indeed
brown dwarfs (Fig.\,\ref{Li6708_spec}).  While the presence of Li is
most common in brown dwarfs, a few older late-M and L dwarfs have been
detected which do not show Li, including DENIS-P J1208+0149 (M9) and
DENIS-P J0090-0658 (L0), Mart\'\i n et al. (\cite{martin99}).

\begin{figure}[h]
\includegraphics[width=0.35\textwidth, angle=270]{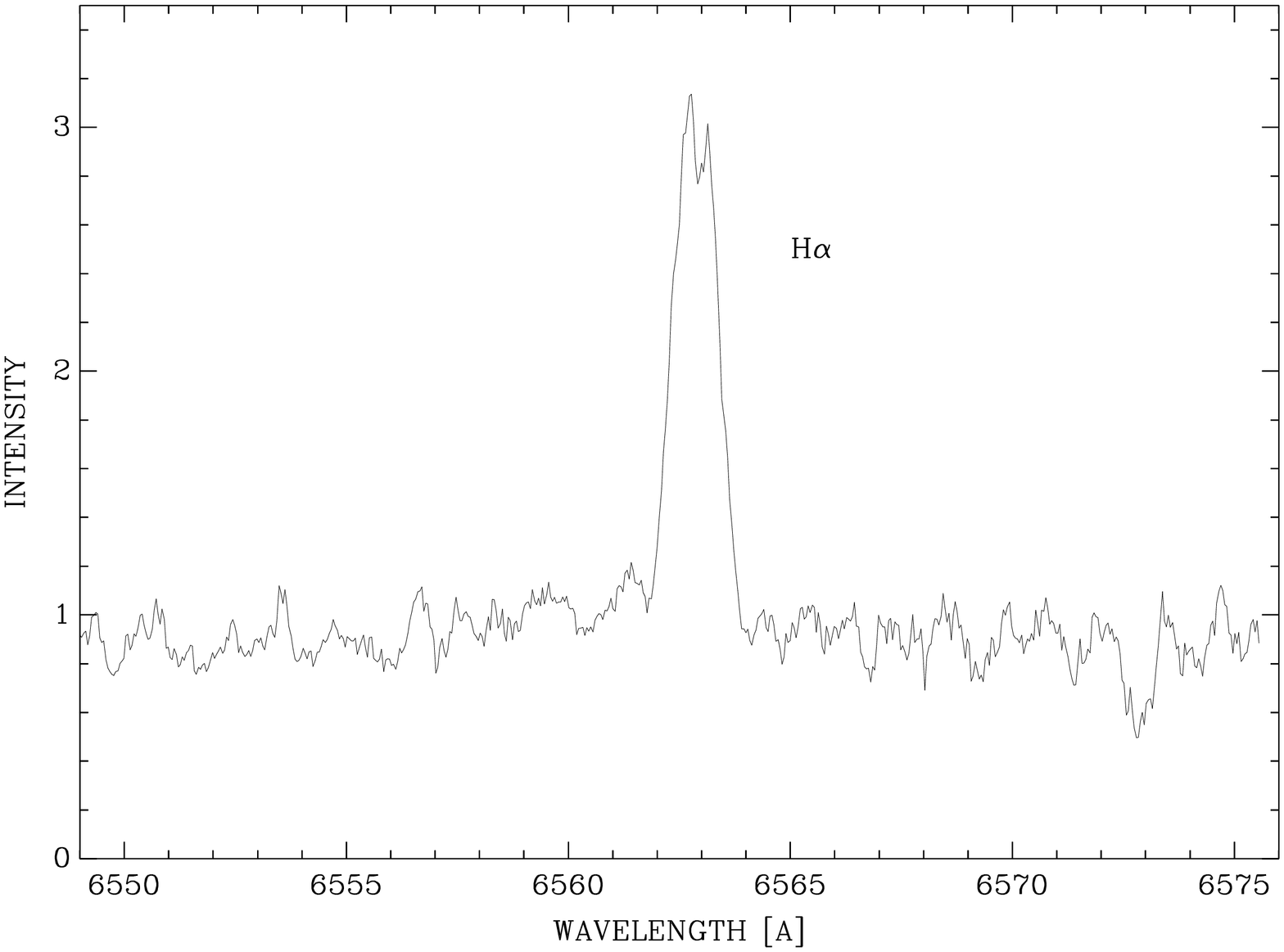}
\caption{The figure shows the H$\alpha$-emission line
of HIP\,16548\,B,C.}
\label{Halpha}
\end{figure}

\begin{figure}[h]
\includegraphics[width=0.35\textwidth, angle=270]{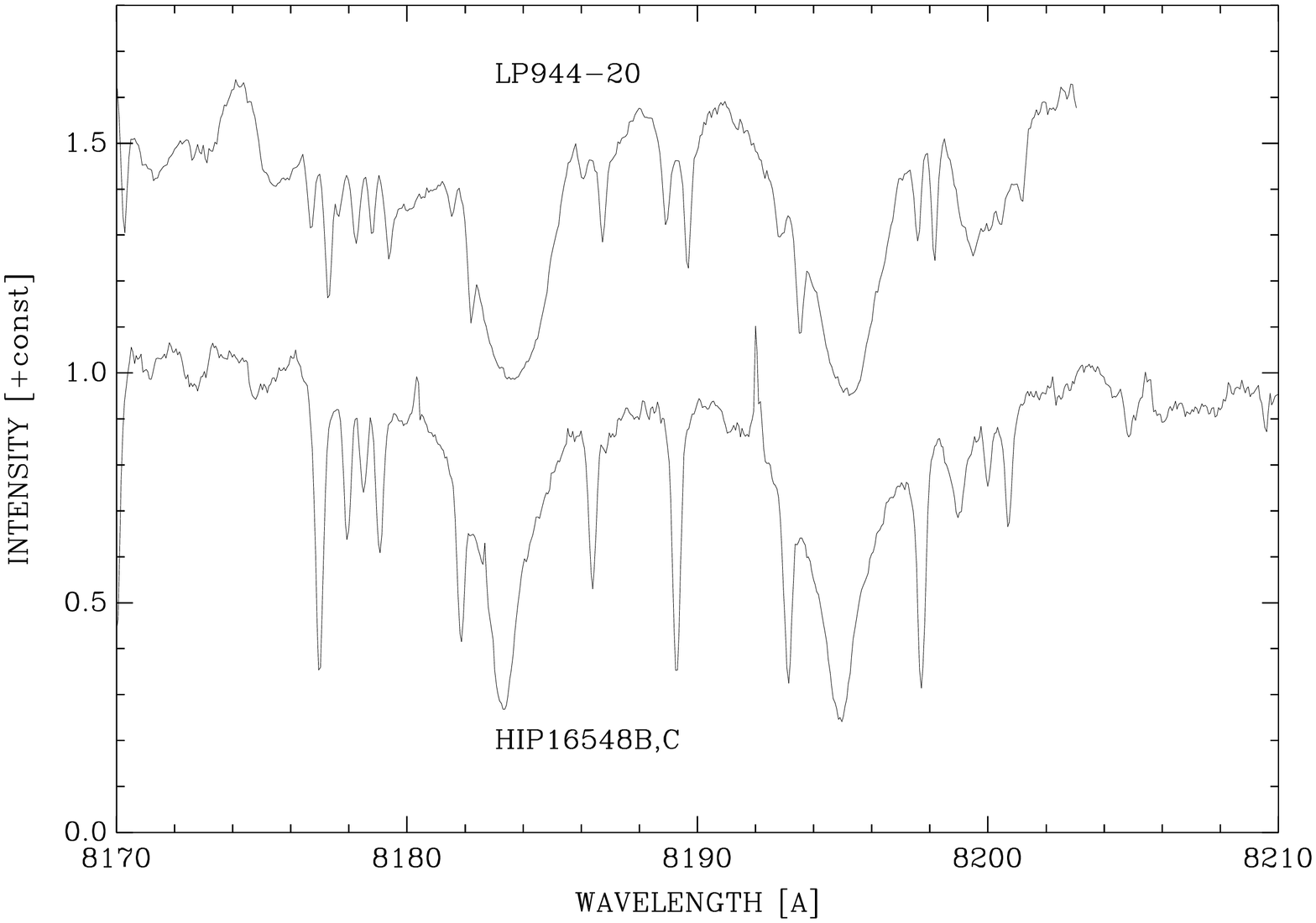}
\caption{The figure shows the NaI doublet at 8183 and 8195 \AA and for
comparison in the spectrum of the brown dwarf LP944-20 is shown below.}
\label{NaI8183_spec}
\end{figure}

\begin{figure}[h]
 \includegraphics[width=0.35\textwidth, angle=270]{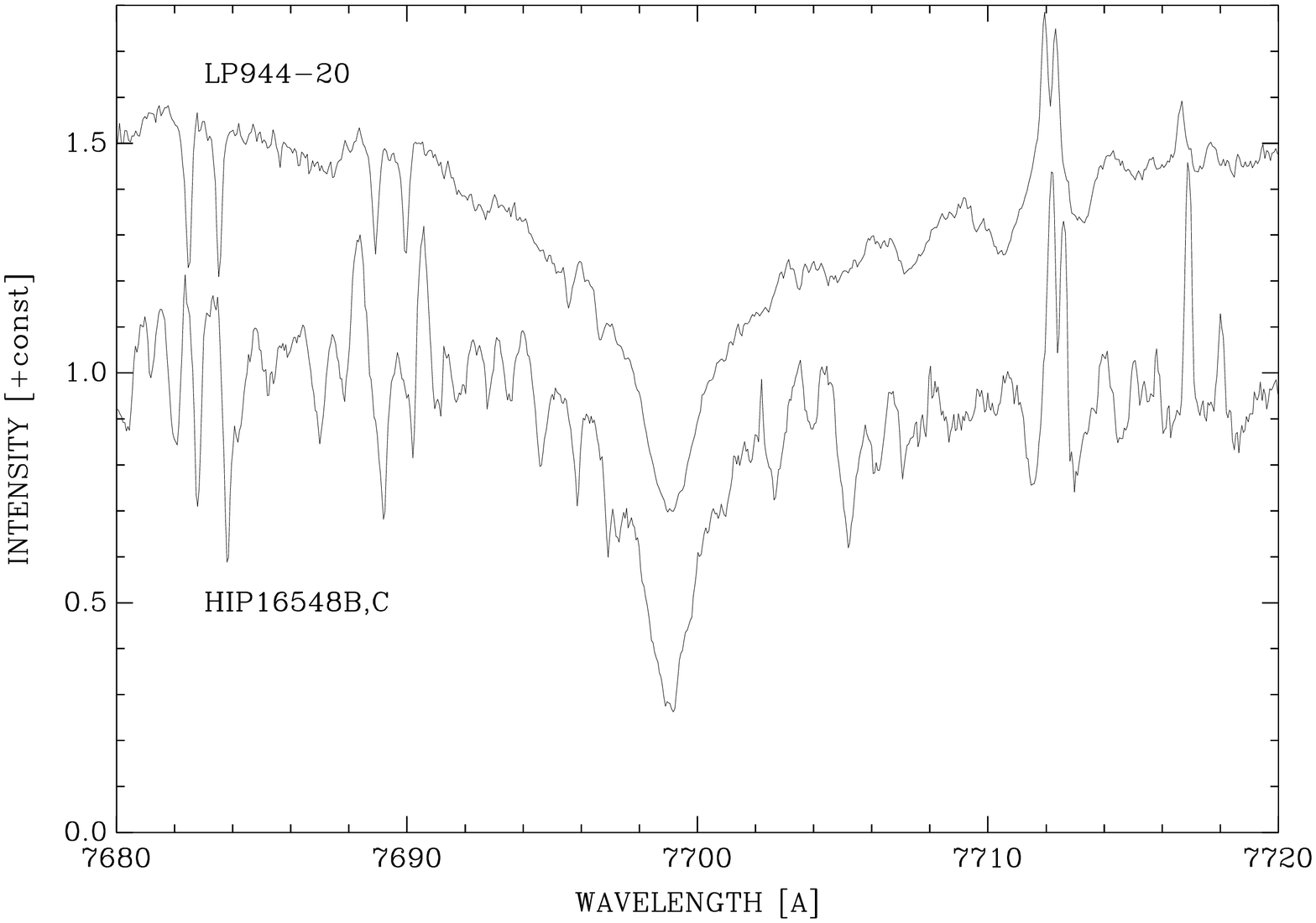}
 \caption{The figure shows the KI 7699 line in the 
 spectrum of HIP\,16548\,B,C and for comparison in the
 spectrum of the brown dwarf LP944-20 which has about the
 same age and the same spectral type as HIP\,16548\,B,C.}
 \label{KI7699_spec}
 \end{figure}

\begin{figure}[h]
 \includegraphics[width=0.35\textwidth, angle=270]{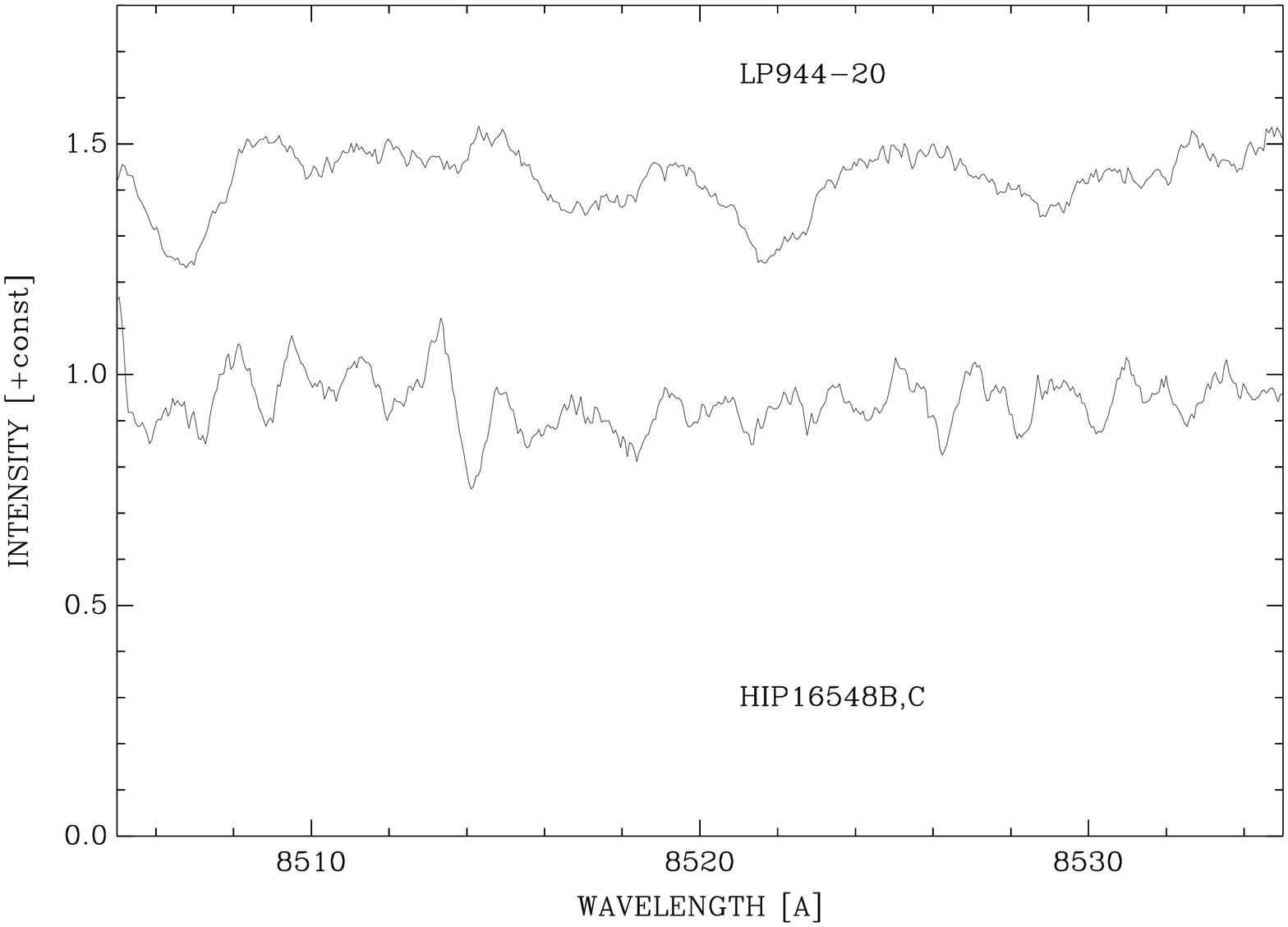}
 \caption{The figure shows the region around the CsI 8521 \AA -line for
 HIP\,16548\,B,C and LP944-20. This line is not seen in the spectrum of
 HIP\,16548\,B,C.}
 \label{CsI8521_spec}
 \end{figure}

The spectra of \object{HIP\,16548\,B} and C shows $H\alpha$ in emission.
The equivalent width is -2.3 \AA , which is fairly typical for such
late-type objects (Fig.\,\ref{Halpha}). The strongest atomic lines are
the NaI doublet at 8183 and 8195 \AA\,(Fig.\,\ref{NaI8183_spec}) and the
KI resonance doublet at 7665 and 7699 \AA\,(Fig.\,\ref{KI7699_spec}),
which are not only clearly visible in the spectrum of
\object{HIP\,16548\,B,C} but also have roughly the same equivalent width
as of LP944-20. Another characteristic feature of young brown dwarfs is
the CsI 8521 \AA -line (Fig.\,\ref{CsI8521_spec}).  This line is however
not seen in the spectrum of \object{HIP\,16548\,B,C}.

\begin{table*}
\caption{Results}
\begin{tabular}{llcccccccc}
\hline \hline
            & type & prim$^5$ & prim$^5$ & prim$^5$ & PA & separation & secon.$^6$ & secon.$^6$ & secon.$^6$ \\
            &      & spec     & $m_J$    & $m_{K'}$ &    & [arcsec]   & $m_J$   & $m_{K'}$ & spec \\
\hline
HD\,28099   & binary      & G2V   &      & 6.6 & $191.5\pm0.5^o$ & $0.43\pm0.01$   & & $11.47\pm0.09$  & M7V-M8V \\
HD\,30505   & binary      &       & 8.2  & 7.1 & $107.2\pm0.4^o$ & $0.408\pm0.005$ & & $11.97\pm0.10$  & M7V\\
HIP\,16908  & binary      & K2V   & 7.8  &     & $266.2\pm1.0^o$ & $0.77\pm0.02$   & $10.4\pm0.1$ & & M1V-M3V \\ 
HD\,286589  & binary      & K5V   & 8.7  &     & $247.8\pm0.5^o$ & $0.62\pm0.02$   & $12.1\pm0.1$ & & M5V-M6V \\ 
            &             &       &      & 7.7 & $248.5\pm0.5^o$ & $0.63\pm0.07$   & & $10.48\pm0.10$ & M5V-M6V \\
LP\,415-176 & binary      & K2V   & 10.0 &     & $322.4\pm0.1^o$ & $4.26\pm0.03$   & $11.9\pm0.1$ & & M7V-M8V \\
HIP\,16548  & triple      & M0V   & 9.2  &     & $268.1\pm0.1^o$ & $2.77\pm0.02$   & $13.8\pm0.3$ & & M8V-M9V \\
            &             &       &      &     & $274.3\pm0.1^o$ & $2.78\pm0.02$   & $14.0\pm0.3$ & & M8V-M9V \\
V1102\,Tau  & binary      & M1V   & 9.2  &     & $78.6\pm0.5^o$  & $1.60\pm0.03$   & $10.5\pm0.3$ & & M2V-M4V \\
HAN\,513    & binary      & M0.5V & 9.3  &     & $25.9\pm1.3^o$  & $0.52\pm0.02$   & $11.0\pm0.1$ & & M2V-M4V \\
HAN\,172    & binary$^4$  & M1V   & 9.5  &     & $106.0\pm0.8^o$ & $0.83\pm0.02$   & $13.4\pm0.5$ & & M5V-M9V \\
HIP\,15720  & binary?$^1$ & K7V   & 6.3  & & & & & \\
BD+08\,642  & binary?$^2$ & K5V   & 8.0  & & & & & \\
BD+16\,630  & binary?$^3$ &       & 10.0 & & & & & \\
J\,332      & active$^4$  & K5V   & 9.4  & & & & & \\  
LP\,415-378 & active      & M0V   & 9.6  & & & & & \\
\hline\hline
\end{tabular}
\label{tab:results}
\\
$^1$ the stellar image is a little elongated. The stellar image is 0.120x0.140 arcsec, the
elongation has a PA of $210\pm10^\circ$. \\
$^2$ the stellar image is a little elongated. The stellar image is 0.113x0.157 arcsec, the
elongation has a PA of $165\pm10^\circ$. \\
$^3$ this star shows a relatively large amplitude of the RV-variation and thus could be
a short period binary. \\
$^4$ this image was taken under bad conditions, limiting the accuracy of the
photometry \\
$^5$ value for primary
$^6$ value for secondary
\end{table*}



\subsection{The other companion candidates}

Table\,\ref{tab:results} gives an overview of the values obtained for
the stars. \object{HD\,28099} is a G2V star. The companion has a
$K'$ magnitude of $11.97\pm0.10$ which implies that it is a M7V to M8V
star.

\object{HD\,30505} was observed with both instruments. Because the
separation is $0.408\pm0.005$ arcsec the secondary was behind the
coronographic mask in the NACO observations. Although NACO image is
slightly elongated, we prefer to use the Keck data only, which were
taken without the mask.  The companion has $K'$ magnitude of
$11.47\pm0.09$, which implies that the companion is an M7V star.

\object{HIP\,16908} is a K2V star, and its companion has an $M_{\rm
J}=7.1\pm0.3$, and thus a M1V to M3V-star. This binary was first
detected by Patience et al. (\cite{patience98}) in a speckle
interferometry survey of the Hyades. These authors estimate the
companion equal to 0.3 $M_\odot$, which is in good agreement with our
measurement. No other companion was found.  We searched for a possible
3rd companion, but none was found in our observations.

\object{HD\,286589} is another star which we observed with both
instruments. Fig.\,\ref{HD286589-NaCo} shows the {\it J-band} image and
Fig.\,\ref{HD286589b-keck} the $K'$-band image. North is up and East is
left in both images. From the images we obtain position angles of
$247.8\pm0.5^o$ and $248.5\pm0.4^o$, and separations of $0.62\pm0.02$
and $0.63\pm0.07$ arcsec, respectively. The agreement between the two
data-sets is thus excellent.  We determine a {\it J-band} magnitude of
$12.1\pm0.1$ mag, and a $K'$-band magnitude of $10.48\pm0.10$ mag. We
thus conclude that the K5V star \object{HD\,286589} is orbited by a M5V
to M6V star. The projected separation currently is 29 AU.

The K2V star \object{LP\,415-176} is also a binary. The separation
is $4.26\pm0.03$ arcsec. The {\it J-band} magnitude of the companion is
$11.9\pm0.1$ mag which implies that it is a M7V to M8V star.  Our aim
was to searched for additional components but did not find any.
\object{V1102\,Tau} is an M1V star. In this case, we found a
companion-candidate with $M_{\rm J}=7.2\pm0.3$ mag, which corresponds to
an M2V to M4V star.  \object{HAN\,513} is an M0.5V-star with a M2 to M4V
companion.

\object{HAN\,172} is an M1V star, which was observed in less than ideal
conditions.  Nevertheless, we found one stellar companion-candidate with
an absolute J-magnitude of $10.1\pm0.5$ mag. The companion is presumably
an M5V-M9V star, and has a mass between 0.08 to 0.12 $M_{\odot}$
(Chabrier et al. \cite{chabrier00}). Unfortunately the star was observed
with the mask when the conditions were less than ideal so that the
accuracy of the photometry is rather limited.  The mass of the object is
close to the brown dwarf boarder but the object probably is not a brown
dwarf. Usually, an object with a mass higher than 0.075 $M_\odot$ is
considered a star, and an object with a mass lower than 0.070 $M\odot$ a
brown dwarf but the exact position of the boundary depends on
metallicity (Chabrier et al. \cite{chabrier00}).  This means that at the
age of the Hyades, the boundary is somewhere around M8V, or M9V. The
brown dwarf LP944-20 which has about the same age as the Hyades has a
spectral type of M9V. The mass of it is between 0.056 and 0.064
$M_{\odot}$ (Tinney \& Reid \cite{tinney98}).  DENIS-P J144137.3-094559
which recently turned out to be member of the super Hyades has a mass of
$0.072^{+0.026}_{-0.022}$ $M_\odot$ and a spectral type of L1V (Seifahrt
et al. \cite{seifahrt04}).

\subsection{Objects without companion candidates}

\begin{figure}[h]
\includegraphics[width=0.40\textwidth, angle=270]{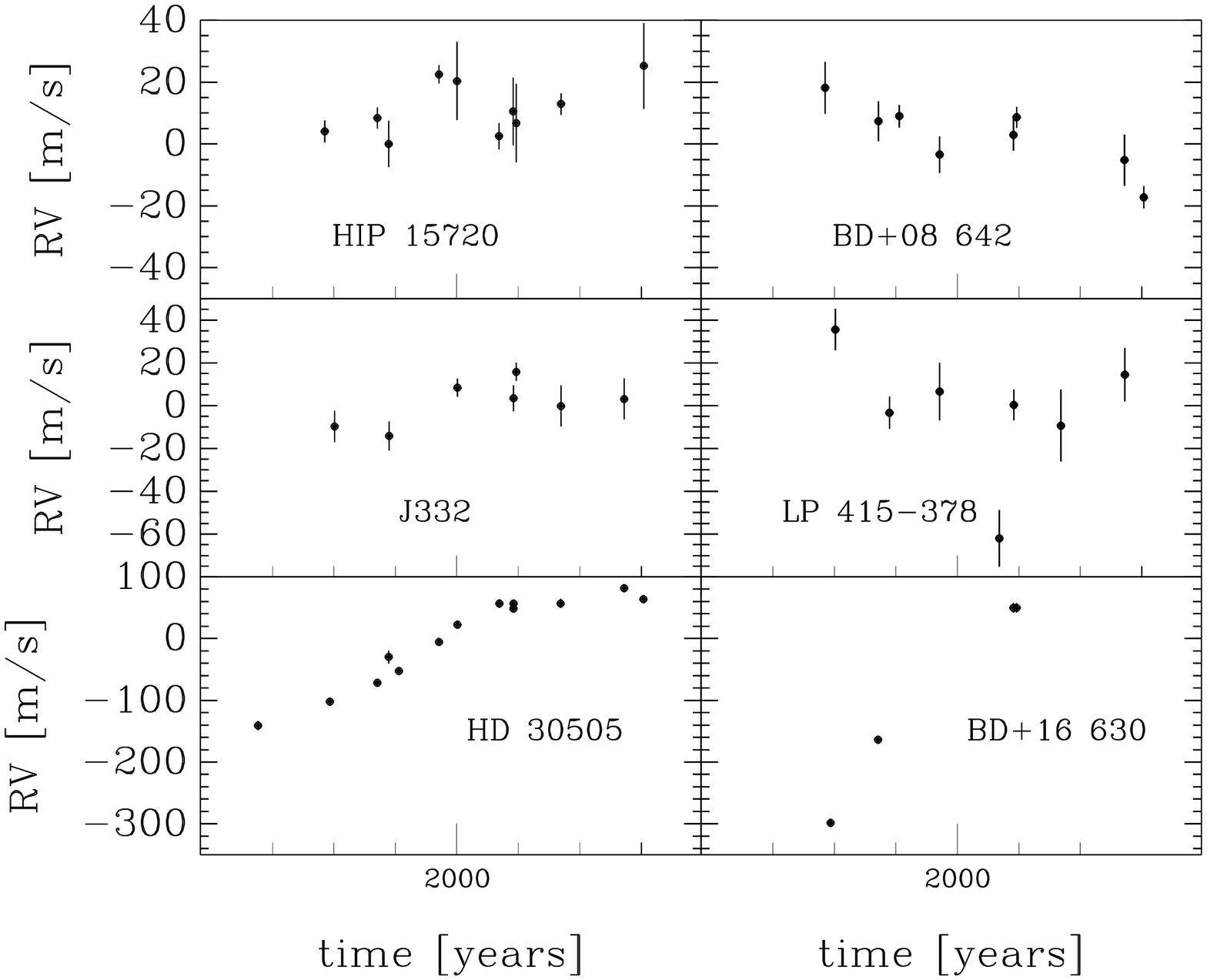}
\caption{This figure shows the RV measurements obtained for those stars,
were we did not find obvious companion candidates in the direct images,
and HD\,30505 for comparison.  In the case of \object{BD+08\,642} and
\object{HIP\,15720} the stellar images are slightly elongated, which may
imply that they are binaries.  The fact that the RV-amplitude of the
three data-points of \object{BD+16\,630} are large implies that this object
probably is a short-period binary.  In the case of \object{J332} and
\object{LP\,415-378}, we argue that the RV-variations are caused by stellar
activity.  }
\label{rv_valuesII}
\end{figure}

Fig.\,\ref{rv_valuesII} shows the RV measurements for those stars, where
we did not find obvious companion candidates (\object{BD+16\,630},
\object{LP\,415-378}, \object{HIP\,15720}, and \object{BD+08\,642}).  
\object{BD+16\,630}, \object{HIP\,15720}, and \object{BD+08\,642}
were however observed with the semi-transparent coronographic mask so
that we can not fully exclude the presence of a faint companion with a
separation of less than 0.35 arcsec.

The stellar images of \object{HIP\,15720} and \object{BD+08\,642} are
slightly elongated. Because the coronographic mask is
semi-transparent, the images of the stars are visible, and not
saturated.  In the case of \object{HIP\,15720} the image is 17\% longer
in one direction than in the other, and in the case of
\object{BD+08\,642} the difference is 40 \%. These stars thus might also
be binaries.  While the elongation of the images is larger than the
typical $\leq$ 10\% elongation of the other stellar images, only further
observations will show whether also these objects are binaries or not.

\begin{figure}[h]
\includegraphics[width=0.35\textwidth, angle=270]{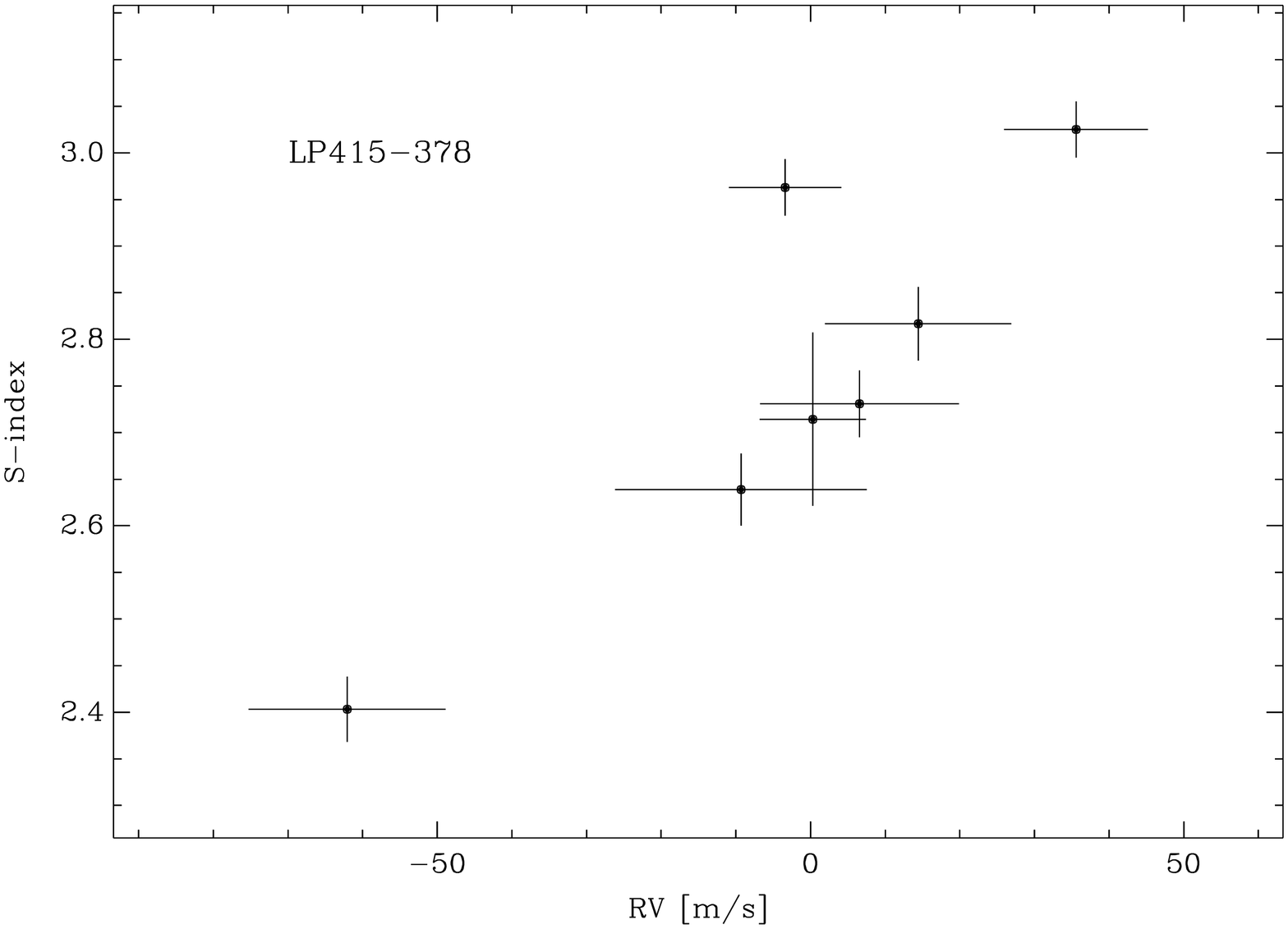}
\caption{The figure shows the S-index versus the RV for LP\,415-378. The
correlation between the two indicates that the RV-variations are
probably caused by stellar activity.}
\label{LP415-378_S-index}
\end{figure}

No indications for binarity were found in the image for \object{J332},
\object{LP\,415-378}, and \object{BD+16\,630}. As mentioned above,
\object{J332} was observed in less than ideal conditions.  The
RV-variations of \object{J332} and \object{LP\,415-378} are most likely
caused by stellar activity.  The low level of the variations in
\object{J332} are consistent with previous results (Paulson et
al. \cite{paulson02}). Fig.\,\ref{LP415-378_S-index} shows the S-index
(Paulson et al. \cite{paulson02}) versus the RV for
\object{LP\,415-378}. There is a clear trend of activity with RV which
indicates that the large amplitude in RV for this star is most likely
caused by stellar activity.

In the case of \object{BD+16\,630} we have only three RV measurements.
However, the relatively large amplitude implies that this object probably
is a short-period binary, and the separation of two components is
simply to small.

\section{Conclusions}

The aim of the project is to search for brown dwarf companions orbiting
Hyades stars within 8 AU using precise RV measurements as input
information.  If the distance would be $\leq$ 3 AU we would have
observed a full orbit, and semi-amplitude of the RV-variations would be
larger than 200 m/s. Such an object would be easily detectable in the RV
data, and we can rule out such an object for all 98 Hyades stars
observed.  Brown dwarfs orbiting at a distance between 3 AU and 8 AU,
would show up as trends of the RV. The problem, however is that such
trends of the RV can also be caused by stellar companions.  The
idea of the project is search for the stellar companions which can
easily be detected by means of AO-imaging, because they are relatively
bright, and have a separation $\geq$ 0.5 arcsec.  The detection of a
stellar companion thus implies that RV trend is caused by this companion
and not by a brown dwarf. An object with a low-amplitude RV trend,
without a stellar companion would then be a brown dwarf companion
candidate.

From an RV survey of 98 stars, we have selected all the stars that could
possibly have a long-term trends in RV, 14 stars in total. In nine of
these we found companions. One of these stars has two companions with
masses between 0.07 to 0.08 $M_{\odot}$. These objects are thus close to
the boarder to the brown dwarf regime, though without the presence of Li
and Cs, we can not confirm that they are indeed brown dwarfs.  Another
star is orbited by a companion with a mass between 0.08 to 0.10
$M_{\odot}$.  In this respect it is interesting to note that the dM4.5e
star G124-62 which is orbited by two L-dwarfs of
$0.072^{+0.026}_{-0.022}$ $M_{\odot}$ recently turned out to be a member
of the super Hyades (Seifahrt et al. \cite{seifahrt04}). From the
remaining five stars, two show evidence that the RV-variations are
caused by stellar activity, and one star seems to be a short-period
binary. Thus there are only two stars, which might have sub-stellar
companions. However, the NACO images of these stars are elongated which
might imply that also these could be binaries.

From the RV-data alone, we can exclude that any of the 98 Hyades
stars studied has a brown dwarf companion with distance of $\leq$ 3 AU.
From the 14 stars selected, at least 9 and possibly 12 have companions
with masses $\geq$ 0.07 $M\odot$, and the two remaining ones are
just active. We did find three objects which are close to the boarder
between stars and brown dwarfs. Assuming that the stars with elongated
images are also binaries, we can also exclude brown dwarf companions
orbiting at a distance between 3 and 8 AU for all stars observed.
We thus estimate the number of companions with masses between 10 $M_J$
and 70 $M_J$ with separations between $\leq$ 8 AU in the Hyades as
$\leq$ 2\%.  We found three companions with masses between 0.08 to 0.10
$M_{\odot}$ at separations of 130 and 40 AU. The frequency of such
objects in the Hyades thus is about $3\pm2$ \%.

As mentioned above there is a general agreement that the frequency of
brown dwarfs orbiting old stars in the solar neighbourhood at distances
of $\leq 3$~AU is about $\leq 0.5\%$. In here we have first of all show
that there is also a brown dwarf desert for stars in the Hyades which
are much younger and have a higher metallicity than the stars in the
solar neighbourhood, and secondly we show that the brown dwarf desert
extends outwards at least up to 8 AU. Whether the brown dwarf desert
extends even further outwards can not be said on the basis of this
survey.


\begin{acknowledgements}

We are grateful to the user support group of the VLT.  Some of the data
presented in this paper were obtained at the W. M. Keck Observatory,
which is operated as a scientific partnership among the California
Institute of Technology, the University of California, and the National
Aeronautics and Space Administration. The Observatory was made possible
by the generous financial support of the W. M. Keck Foundation. This
work made use of the SIMBAD database operated by CDS, France, and data
products from the Two Micron All Sky Survey, which is a joint project of
the University of Massachusetts and the Infrared Processing and Analysis
Center/California Institute of Technology, funded by the National
Aeronautics and Space Administration and the National Science
Foundation.  Funding for JP's fellowship was provided to the Michelson
Science Center under its Michelson Fellowship Program.  Portions of this
material is based upon work supported by the National Aeronautics and
Space Administration under Grant NNG04G141G issued through the
Terrestrial Planet Finder Foundation Science program, as well as under
the auspices of the U.S. Department of Energy by the University of
California, Lawrence Livermore National Laboratory under Contract
W-7405-ENG-48, and also supported in part by the National Science
Foundation Science and Technology Center for Adaptive Optics, managed by
the University of California at Santa Cruz under cooperative agreement
No. AST 9876783.

\end{acknowledgements}

\end{document}